\documentclass[final,english]{bullsrsl}[2022/06/15]
\newcommand{\specialcell}[2][c]{%
\begin{tabular}[#1]{@{}c@{}}#2\end{tabular}}

\usepackage[latin1]{inputenc}
\usepackage[T1]{fontenc}

\usepackage{natbib} 
\usepackage{graphicx}

\begin{document}
\title{Observation of mulitply imaged quasars with the 4-m ILMT}

\author[affil={1,2}, corresponding]{Talat}{Akhunov}
\author[affil={3,4}]{Bhavya}{Ailawadhi}
\author[affil={5}]{Ermanno}{Borra}
\author[affil={3,6}]{Monalisa}{Dubey}
\author[affil={3,6}]{Naveen}{Dukiya}
\author[affil={7}]{Jiuyang}{Fu}
\author[affil={7}]{Baldeep}{Grewal}
\author[affil={7}]{Paul}{Hickson}
\author[affil={3}]{Brajesh}{Kumar}
\author[affil={3}]{Kuntal}{Misra}
\author[affil={3,4}]{Vibhore}{Negi}
\author[affil={8}]{Anna}{Pospieszalska-Surdej}
\author[affil={3,9}]{Kumar}{Pranshu}
\author[affil={7}]{Ethen}{Sun}
\author[affil={8}]{Jean}{Surdej}

\affiliation[1]{National University of Uzbekistan, Department of Astronomy and Astrophysics, 100174 Tashkent, Uzbekistan}
\affiliation[2]{ Ulugh Beg Astronomical Institute of the Uzbek Academy of Sciences, Astronomicheskaya 33, 100052 Tashkent, Uzbekistan}
\affiliation[3]{Aryabhatta Research Institute of Observational sciencES (ARIES), Manora Peak, Nainital, 263001, India}
\affiliation[4]{Department of Physics, Deen Dayal Upadhyaya Gorakhpur University, Gorakhpur, 273009, India}
\affiliation[5]{Department of Physics, Universit\'{e} Laval, 2325, rue de l'Universit\'{e}, Qu\'{e}bec, G1V 0A6, Canada}
\affiliation[6]{Department of Applied Physics, Mahatma Jyotiba Phule Rohilkhand University, Bareilly, 243006, India}
\affiliation[7]{Department of Physics and Astronomy, University of British Columbia, 6224 Agricultural Road, Vancouver, BC V6T 1Z1, Canada}
\affiliation[8]{Institute of Astrophysics and Geophysics, University of Li\`{e}ge, All\'{e}e du 6 Ao$\hat{\rm u}$t 19c, 4000, Li\`{e}ge, Belgium}
\affiliation[9]{Department of Applied Optics and Photonics, University of Calcutta, Kolkata, 700106, India}

\correspondance{talat77@rambler.ru; akhunovtalat@gmail.com}
\date{27th May 2023}
\maketitle


%

\begin{abstract}
Gravitationally lensed quasars (GLQs) are known to potentially provide an independent way of determining the value of the Hubble-Lema\^{i}tre parameter $H_{0}$, to probe the dark matter content of lensing galaxies and to resolve tiny structures in distant active galactic nuclei.

That is why multiply imaged quasars are one of the main drivers for a photometric monitoring with the 4-m International Liquid Mirror Telescope (ILMT). We would like to answer the following questions -- how many multiply imaged quasars should we be able to detect with the ILMT? And how to derive accurate magnitudes of the GLQ images?  Our estimation of the possible number of multiply imaged quasars is $15$, although optimistic forecasts predict up to $50$ of them. We propose to use the adaptive PSF fitting method for accurate flux measurements of the lensed images. During preliminary observations in spring 2022 we were able to detect the quadruply imaged quasar - SDSS J1251+2935 in the $\it{i}$ and $\it{r}$ spectral bands. 
\end{abstract}


\section{Introduction}
The great interest in gravitational lensing comes from the fact that this phenomenon can be used as an astrophysical and cosmological tool (cf. \cite{Surdej2001,Claescens2002}). Analysis of multiply imaged quasars (which undergo strong gravitational lensing) may reveal the structure and composition of tiny regions in distant active galactic nuclei, of intervening galaxy haloes, and of intergalactic space. So gravitational lensing may enrich our view of the distant Universe and affect our physical understanding of various classes of extragalactic objects (e.g. \cite{Schneider1993,Schneider2006}). 

Observations with the ILMT will provide continuous and homogeneous sets of observational data in the $\it g$, $\it r$ and $\it i$ spectral bands \citep{Surdej2018}. The main goals of a photometric monitoring is the determination of time delays between the multiple lensed images of quasars and the observation of microlensing events. At the same time, we are also interested in questions such as -- how many lensed quasars should we be able to observe with the ILMT and which photometric techniques should we use for flux measurements of very nearby lensed QSO images? Moreover, preliminary observations in May-June 2022 made it possible to detect at least one known case of multiply imaged quasar: SDSS J1251+2935 in the $\it{r}$ and $\it{i}$ spectral bands.

In this article, we give an estimation of the possible number of GLQs that may be detected during photometric observations, propose photometric measurement methods based on image subtraction techniques, and estimate the total flux of the observed multiply imaged quasar SDSS J1251+2935.

\section{Probable Number of GLQs in the FOV of the ILMT}
First estimation of the probable number of quasars and GLQs which can be detected in the field of view of the ILMT was provided by \cite{Surdej2001,Claescens2002}. The optical depth $\tau_{q}$ for the formation of multiply imaged quasars at redshift $z_{q}$ by galaxies distributed along their line-of-sight is given by the relation:

\begin{equation}
\tau_{q} = \frac{4}{15} F \bigg(1-\big(1+z_{q}\big)^{-0.5}\bigg)^3 
\end{equation}

where the parameter $F$ is the effectiveness of galaxies to produce multiply imaged quasars. So, adopting $F \cong 0.047$ \citep{Fukugita1996}, we  find $\tau_{q} \cong 3.15 \times 10^{-4}$, $ 6.22 \times 10^{-4}$ and $ 9.46 \times 10^{-4}$  for $z_{q} = 1$, $1.5$ and $2$, respectively. The total depth for all redshifts was estimated to be $\tau_{q} \cong 2.5 \times 10^{-3}$. At that time, it was assumed that the ILMT would be able to detect about $20,000$ quasars and that about 50 of them could be composed of multiple lensed images.

Later these values were re-estimated by \cite{Finet2013}. The probable number of quasars was estimated as $\sim$ 9072 and about 22 of them to be GLQs. However, more recently, by comparing and cross-correlating various catalogues (Milliquas, Gaia-DR2, etc.), a new estimation of the number of quasars that will fall into the FOV of the ILMT has been derived by \cite{Mandal2020}. The final quasar catalogue for the ILMT contains 6738 objects, and accordingly the probable number of lensed quasars is about 15. The redshifts of the majority of the lensed quasars will be in the range $z_{q}=[0.5-2.5]$, and their apparent magnitude $G=[18-22.5]$ mag. 

\section{Data processing and reduction techniques}

After identifying the GLQs, we will determine the PSF from observed point-like sources and then the flux ratios between the lensed components. To quickly determine the positions and magnitudes of sources over the field of view, the Source Extractor software is usually used \citep{Bertin1996}. To cross-check these values, we use other traditional image subtraction methods. For example, in the methods by \cite{Vakulik2004,Stetson1987} they propose to make use of Gauss, Moffat, Lorentz, etc. analytical functions to represent the PSF. This methods strictly assumes a smooth, normal distribution of the fluxes in the point-like sources, which is not always the case. 

For example, for the case of asymmetric or complex stellar images, we previously proposed a highly performing photometric method: the adaptive PSF-fitting technique \citep{Akhunov2017}. In this method the central fluxes of the point sources and unknown PSF can be determined iteratively by solving a set of linear equations. The only requirements specific to the adaptive PSF-fitting method are: (i) the positions of each source on the CCD frame have to be accurately known, at least within a fraction of the pixel size; (ii) the CCD frames must have a good signal-to-noise ratio.

This method only works with a numerical PSF and does not require an analytical representation. Secondly, there is no need for the assumption of PSF invariance over large regions of the CCD frame. To calculate the fluxes of multiple lensed images, we assume PSF invariance over only the object area. The problem of deriving the fluxes of the lensed QSO images is reduced to solving a system of ordinary linear equations. The shape and width of the light distribution of the PSF do not really matter. Our method can also be successfully used for defocused images.

\begin{figure}[t]
\centering
\includegraphics[width=0.5\textwidth]{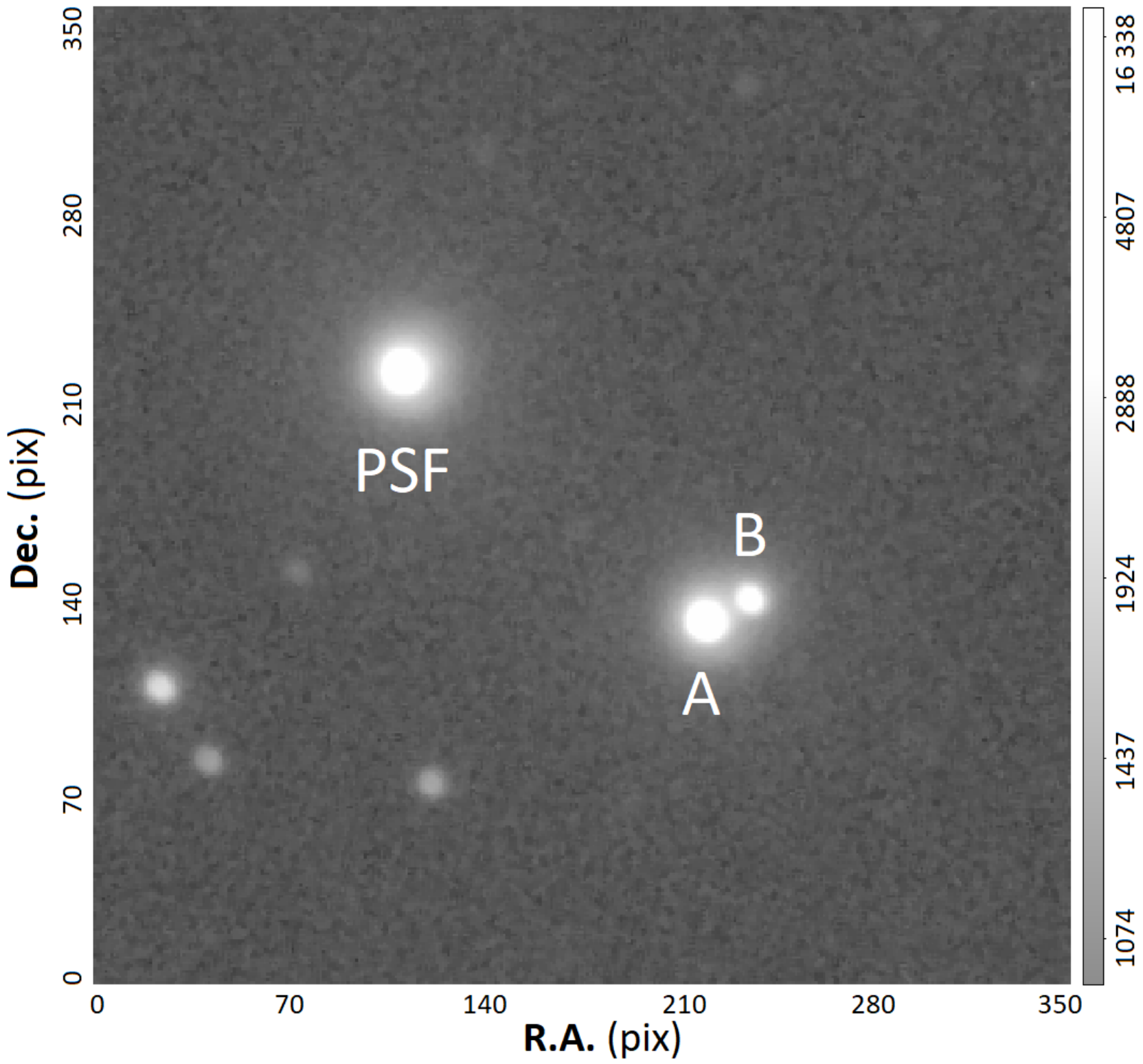}
\bigskip
\begin{minipage}{12cm}
\caption{Image of an optical pair, considered as a possible case of doubly imaged quasar (noted as A and B) and a star (noted as PSF) which we use to calculate the analytical PSF. North is up and East to the left. The data is taken from the ILMT CCD frame \texttt{20221026-060.fits} ($\it {g}$ filter) and is displayed in ADU.}
\label{01_fig}
\end{minipage}
\end{figure}

Let us compare the photometric outputs of these methods by applying them to some observational data that simulate the case of a GLQ composed of two lensed images. This is an ordinary optical pair at the 2000 equatorial coordinates: 106.50509, +29.28322 (see Fig.\,\ref{01_fig}). Here we are interested in the magnitude difference between the $A$ and $B$ components $\Delta m = m_{A} - m_{B}$. The Gaia database reports the value $\Delta m_G = -2.173\pm0.004$ mag. Estimations of the magnitude difference using different photometric methods are given in Table\,\ref{obs_data}. First, we made use of the Source Extractor software which does only report rough magnitude estimates for the case of nearby components. 

\begin{figure}[t]
\centering
\includegraphics[width=0.875\textwidth]{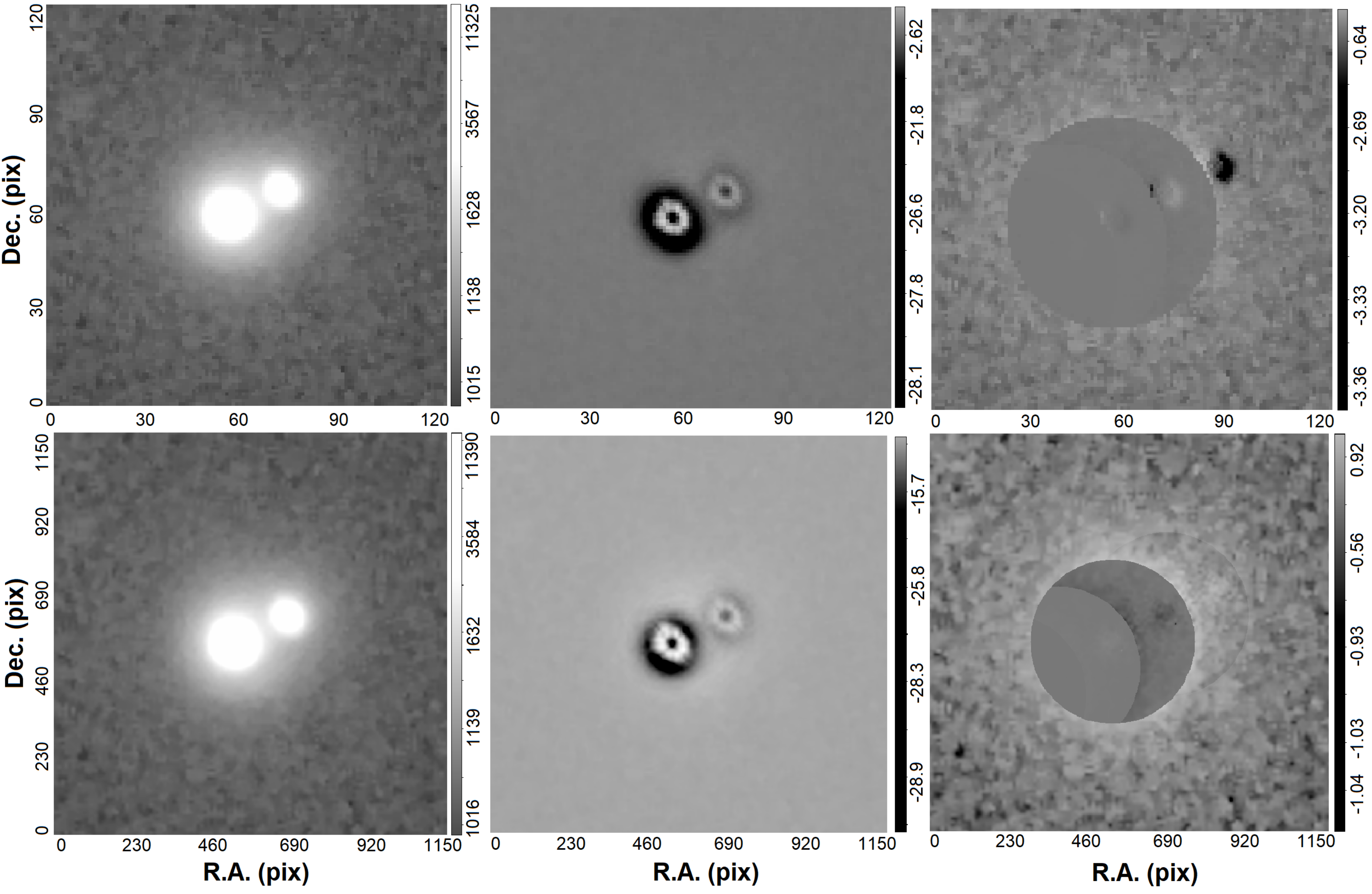}
\bigskip
\begin{minipage}{12cm}
\caption{Top: Zoomed image of the stellar pair, subtractions of two Moffat PSFs and adaptive PSF methoda are shown. Bottom: The same while making use of the smoothed images are presented. Original and smoothed frames are displayed in ADU, and the residual images are shown in units of the standard noise deviation (see the vertical colour bars on the right sides of the frames).}
\label{02_fig}
\end{minipage}
\end{figure}

We may get more interesting results after applying the image subtraction technique (see Fig.\,\ref{02_fig}). The original image, the residuals after fitting of two Moffat's functions and the adaptive PSF are presented in the top of Fig.\,\ref{02_fig}. At the same time, in order to tackle the case of very nearby components, we smoothed out our original frame using a 10-times pixelation and bi-cubic interpolation technique (see \cite{Akhunov2017} for more details). 

In the bottom of Fig.\,\ref{02_fig} we presented the results of the adaptive PSF-fitting method on to a bi-cubic interpolated CCD frame. On the left side there is a 10 times pixelated image, which means that each pixel of the original CCD frame has been bi-cubic interpolated from the 10 inner points equidistant from each other. The center and right sides illustrate the residuals between the observed and modelled frames (as in the top side).

Somewhat defocused or perturbed stellar images and their brightness distributions cannot be described by simple analytical functions. The difference may reach several dozens of standard noise deviation (see middle frames in Fig.\,\ref{02_fig}). But instead of fitting an analytical PSF our adaptive PSF fitting method leads to acceptable results (see the right side residuals in Fig.\,\ref{02_fig}).

\begin{figure}[t]
\centering
\includegraphics[width=0.75\textwidth]{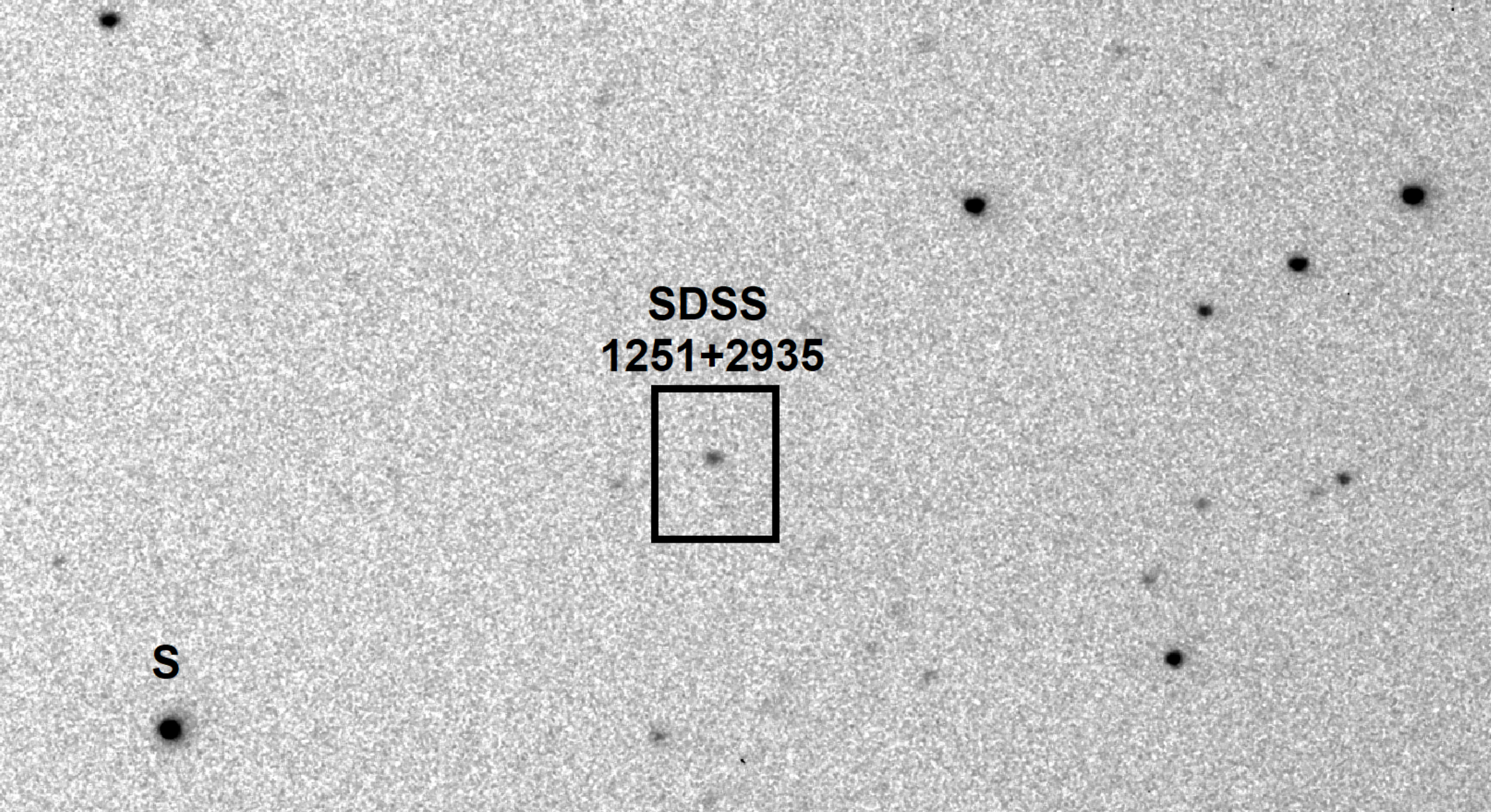}
\bigskip
\begin{minipage}{12cm}
\caption{ILMT composite image of the SDSS J1251+2935 field (8.75 arcmin $\times$ 4.10 arcmin, pixel scale $\sim$0.33 arcsec/pixel). S is a reference star (15.38$\pm$0.01 and 15.21$\pm$0.01 mag in $\it {g}$ and $\it {i}$).}
\label{03_fig}
\end{minipage}
\end{figure}

As we see from Table\,\ref{obs_data} and Fig.\,\ref{02_fig}, reliable results have been obtained using the adaptive PSF fitting method and we propose to use this technique in the future. Details of the application of this method can be found on the cloud: \url{https://drive.google.com/drive/folders/1u3fIKxvZbM2Uz0y2NvcyJGf8BtD-K3sR}.

\begin{table}
\centering
\begin{minipage}{160mm}
\caption{Magnitude differences $\Delta m = m_A-m_B $ for the different methods.}
\end{minipage}
\smallskip

\begin{tabular}{ccccccccc}
\hline
\textbf{Gaia} &~~~& \textbf{Source}    &~~~& \multicolumn{2}{c}{\textbf{Fitting of the}} &~~~& \multicolumn{2}{c}{\textbf{Fitting of the}} \\
              &&    \textbf{Extractor} &&    \multicolumn{2}{c}{\textbf{Moffat PSF}}     &&    \multicolumn{2}{c}{\textbf{adaptive PSF}}   \\
              &&                       &&    \textbf{original} & \textbf{smoothed}       &&    \textbf{original} & \textbf{smoothed}       \\
\hline
\specialcell{-2.173\\$\pm$ 0.004} && \specialcell{-1.061\\$\pm$ 0.004} && \specialcell{-1.972\\$\pm$ 0.002} & \specialcell{-1.951\\$\pm$ 0.002} && \specialcell{-2.182\\$\pm$ 0.002} & \specialcell{-1.986\\$\pm$ 0.001} \\
\hline
\end{tabular}
\label{obs_data}
\end{table}

\section{Images of the GLQ SDSS J1251+2935 detected on the ILMT frames}
During preliminary observations in May-June 2022, we were able to detect images of the GLQ SDSS J1251+2935 on 3 frames taken with the $i$ and $r$ filters. This quasar was discovered by \cite{Kayo2007} and represents a quadruply imaged source with redshift $z_{q}=0.802$. The configuration of the system resembles a crossbow, which is formed after the light rays from the quasar pass the gravity field of a relatively nearby galaxy ($z_{g}=0.410$). The angular separation of component $D$ from the other three ones is approximately 1.79 arcsec, however, the latter are no more than 1.08 arcsec apart from each other, so in ground-based observation they merge into one clot with a total brightness of about $18.9$ mag. The brightness of component $D$ is about $20.2$ mag, which was approximately at that time at the sensitivity limit of the ILMT and detector.

Despite the fact that the nights were partly cloudy during the observations, our detector was able to detect the object in the $g$ and $i$ spectral bands. The ILMT composite image is shown in Fig.\,\ref{03_fig}. Due to non ideal atmospheric conditions, we could not resolve the multiple components, especially the weakest component $D$ did not appear at all. The aperture photometry (\texttt{DAOPHOT}) allowed us to get the total brightness of the $A$, $B$ and $C$ components which is 18.88$\pm$0.17 and 17.80$\pm$0.07 mag in the $g$ and $i$ bands, respectively. Our preliminary observations confirm the possibility of detecting and monitoring photometrically GLQs with the ILMT.

\section{Conclusion}
In this article we considered some aspects of a planned photometric survey of gravitationally lensed quasars using the ILMT. Here we tried to provide some answer to the questions -- how many GLQs will we be able to detect with the ILMT, which techniques shall we use to measure their fluxes and whether the possibility of observing such objects is realistic?
Our estimation of the possible number of GLQs is 15, although optimistic forecasts predict around 50. Next, we proposed several techniques to measure the fluxes of the lensed components. The best technique seems to be the adaptive PSF fitting method.

\begin{acknowledgments}
The 4m International Liquid Mirror Telescope (ILMT) project results from a collaboration between the Institute of Astrophysics and Geophysics (University of Li\`{e}ge, Belgium), the Universities of British Columbia, Laval, Montreal, Toronto, Victoria and York University, and Aryabhatta Research Institute of observational sciencES (ARIES, India). The authors thank Hitesh Kumar, Himanshu Rawat, Khushal Singh and other observing staff for their assistance at the 4m ILMT. The team acknowledges the contributions of ARIES's past and present scientific, engineering and administrative members in the realisation of the ILMT project. JS wishes to thank Service Public Wallonie, F.R.S.-FNRS (Belgium) and the University of Li\`{e}ge, Belgium for funding the construction of the ILMT. PH acknowledges financial support from the Natural Sciences and Engineering Research Council of Canada, RGPIN-2019-04369. PH and JS thank ARIES for hospitality during their visits to Devasthal. B.A. acknowledges the Council of Scientific $\&$ Industrial Research (CSIR) fellowship award (09/948(0005)/2020-EMR-I) for this work. M.D. acknowledges Innovation in Science Pursuit for Inspired Research (INSPIRE) fellowship award (DST/INSPIRE Fellowship/2020/IF200251) for this work. T.A. thanks Ministry of Higher Education, Science and Innovations of Uzbekistan (grant FZ-20200929344).
\end{acknowledgments}

\begin{furtherinformation}

\begin{orcids}
\orcid{0000-0001-5115-6310}{Talat}{Akhunov}
\end{orcids}

\begin{authorcontributions}
This work results from a long-term collaboration to which all authors have made significant contributions.
\end{authorcontributions}

\begin{conflictsofinterest}
The authors declare no conflict of interest.
\end{conflictsofinterest}

\end{furtherinformation}

\bibliographystyle{bullsrsl-en}

\bibliography{S11-P17_AkhunovT}

\end{document}